\documentclass[3p,times,procedia]{elsarticle}

\usepackage{ecrc}

\volume{00}

\firstpage{1}

\journalname{ISTTT Procedia}

\runauth{M.~Treiber, A.~Kesting}

\jid{sbspro}

\jnltitlelogo{Procedia Social and Behavioral Sciences}

\CopyrightLine{2013}{The Authors. Published by Elsevier Ltd. Selection and peer review under responsibility of Delft University of Technology.}

\usepackage{amssymb}

\usepackage[figuresright]{rotating}

\usepackage{units} 

\providecommand{\bc}{\begin{center}}
\providecommand{\ec}{\end{center}}
\providecommand{\be}{\begin{equation}}
\providecommand{\ee}{\end{equation}}
\providecommand{\bea}{\begin{eqnarray}}
\providecommand{\eea}{\end{eqnarray}}
\providecommand{\bdm}{\begin{displaymath}}
\providecommand{\edm}{\end{displaymath}}
\providecommand{\bdma}{\begin{eqnarray*}}
\providecommand{\edma}{\end{eqnarray*}}
\providecommand{\ba}{\begin{eqnarray*}}
\providecommand{\ea}{\end{eqnarray*}}
\providecommand{\bi}{\begin{itemize}}
\providecommand{\ei}{\end{itemize}}
\providecommand{\benum}{\begin{enumerate}}
\providecommand{\eenum}{\end{enumerate}}

\providecommand{\refkl}[1]{(\ref{#1})}

\providecommand{\text}[1]{{\mbox{ #1}}}

\providecommand{\tr}{^{\text{T}}}

\providecommand{\fig}[2]{
   \begin{center}
     \includegraphics[width=#1]{#2}
   \end{center}
}

\providecommand{\abl}[2]{\frac{{\rm d} #1}{{\rm d} #2}}  

\providecommand{\sub}[1]{_{\rm #1}}
\renewcommand{\sup}[1]{^{\rm #1}}

\providecommand{\vecbeta}{\vec{\beta}}
\providecommand{\hatvecbeta}{\hat{\vec{\beta}}}
\providecommand{\hatvec}[1]{\hat{\vec{#1}}}
\providecommand{\tr}{^{\text{T}}} 

\begin{document}

\begin{frontmatter}

\title{Microscopic Calibration and Validation of Car-Following Models
  -- A Systematic Approach}

\author[TUD]{Martin Treiber\corref{cor1}}\ead{treiber@vwi.tu-dresden.de}
\author[TOMTOM,TUD]{Arne Kesting}\ead{mail@akesting.de}

\address[TUD]{Technische Universit\"at Dresden, Institute for Transport \& Economics,\\
W\"urzburger Str. 35, 01062 Dresden, Germany}
\cortext[cor1]{Corresponding author.}

\address[TOMTOM]{TomTom Development GmbH Germany, An den Treptowers 1, 12435 Berlin (Germany).}

\begin{abstract}
Calibration and validation techniques are crucial in
assessing the descriptive and predictive power of car-following models and their
suitability for analyzing traffic flow. 
Using real and generated floating-car and trajectory data,
we systematically investigate following aspects: 
Data requirements and preparation, conceptional approach
including local maximum-likelihood and global LSE
calibration with several objective functions, 
influence of the data sampling rate and measuring
errors,  the effect of data smoothing on the calibration result, and model
performance in terms of fitting quality, robustness, parameter
orthogonality, completeness and plausible parameter values.
\end{abstract}


\end{frontmatter}



\section{\label{sec:intro}Introduction}
Microscopic traffic flow models describe the driving behavior, local
traffic rules, and possible restrictions of the
vehicle. All these aspects vary with the driver but also with the country and with
time. For example, drivers in the United States and in Germany have different
driving styles, drive different types of vehicles and are subject
to different traffic regulations. For drivers in China, the differences are
even more pronounced. Furthermore, traffic flow on a given road is
generally more effective during the
morning rush hours than in the evening rush hours or at night since
 drivers are generally more alert and drive more
effectively in the morning than at other times.
Consequently, not even the best model reproducing all traffic flow
situations in question can be applied to a specific 
task with the default
parameter set. Instead, to
maximize the model's descriptive power, its parameters need to be
estimated based on representative traffic data. 

Nevertheless, only few systematic investigations have been
undertaken on this
topic.
While collective 
macroscopic aspects of traffic flow such as travel times or traffic waves can  be
calibrated and validated on macroscopic
data~\cite{wang2008real,treiber2012validation}, it is more natural to calibrate
car-following models on microscopic data sets. These include single-vehicle data
of stationary detectors, extended floating-car data
(xFCD)~\cite{Brockfeld-benchmark04,Punzo-bench05,Kesting-Calibration-TGF07,Arne-DissOnline,TreiberKesting-Book}, and trajectory
data~\cite{brockfeld2006validating,Kesting-Calibration-TRR08,Ossen-noise-TRR08}.  
Trajectory data such as these of the NGSIM project~\cite{Thiemann-NGSIM-TRB08,punzo2011assessment} contain the dynamics 
of the positions  (and lanes) of \emph{all} vehicles in a given spatiotemporal
region. In contrast, 
xFC data (which typically are generated by instrumented vehicles)
include, for only one or a few vehicles, time
series of the position, the speed, and the gap to the respective
leading vehicle.

Regarding calibration and validation, many questions are not yet settled: What is the
minimum quantitative data requirement for a given calibration task in
terms of the minimal set of observed
quantities, minimum number of
vehicles, minimum length of time interval, or minimum temporal
resolution? Is it possible to formulate qualitative data
requirements by defining a minimal set of traffic states which must be
contained in
the data~\cite{Punzo2009parameters}? To which degree does data noise
or the sampling rate influence the calibration
result~\cite{Ossen-noise-TRR08}?  Is it possible to
distinguish noise from intra-driver and inter-driver
variations~\cite{Ossen-interDriver06,tordeux2010adaptive,ossen2011heterogeneity}?
To which degree does the result
differ when calibrating a given model on given data with different
methods such as least squared errors (LSE), maximum-likelihood, 
or Kalman filtering~\cite{punzo2005part,Wang-RealTimeEstimation-2005,wang2008real}? 
How does the result depend on the objective
function~\cite{Kesting-Calibration-TGF07,Arne-DissOnline}? Why is
there so little difference when comparing LSE calibration 
results of the ``best'' with that of the ``worst''
models~\cite{Brockfeld-benchmark04}? Are there more suitable criteria
than the fitting quality, e.g., sensitivity with respect to parameter
changes~\cite{punzo2011sensitivity} or robustness?

In this contribution, we systematically investigate these aspects
using both real and generated data. While, ultimately, the quality of models and
calibration techniques is to be assessed  by real data, generated data
allow for separating the influencing 
factors in a systematic and controlled way. 

In principle, this program can be performed with any
car-following model specifying the acceleration, either directly
or indirectly in terms of speed differences~\cite{TreiberKesting-Book}. For the sake of
demonstration, we will use mainly the intelligent-driver model
(IDM)~\cite{Opus,TreiberKesting-Book}. It is suited for this kind of investigations since
the meaning of its five parameters is intuitive and each parameter
relates to a certain driving regime: The desired speed $v_0$ is
relevant for
cruising in free-traffic conditions, the desired time gap $T$ pertains
to
steady-state car-following, the minimum gap $s_0$ to creeping and
standing traffic, and the maximum acceleration $a$ and desired
deceleration $b$ relate to non-steady-state traffic flow.
For cross comparison, we will also use the optimal-velocity model
(OVM)\cite{Bando1} and the full-velocity difference
model~\cite{Jiang-vDiff01}. However, we will use these models together
with modified steady-state speed-gap relations
also containing the parameters $s_0$ and $T$ 
instead of the original ones with non-intuitive parameters.

After discussing methodic data, calibration, and simulation issues in the next two
sections, we investigate the influence of the data sampling rate,
noise, serial correlations, and smoothing in Sec.~\ref{sec:smooth}
before discussing the more model-related issues data and model
completeness as well as
parameter orthogonality in Sec.~\ref{sec:complete}. We conclude this
contribution by a discussion 
and hints for further research in Sec.~\ref{sec:concl}.

\section{\label{sec:data}Data Issues}
Since car-following models are supposed not only to describe the
microscopic dynamics of individual drivers but also
macroscopic aspects of traffic flow (traffic waves, traffic jam
propagation, or travel times) emerging from the dynamics, such models
can, in principle, be calibrated and
validated by both microscopic and macroscopic data. Nevertheless, it
is more natural to use microscopic data of individual
drivers. Since this allows to separate the effects of inter-driver
variations (different drivers have different driving styles) from intra-driver
variations (even one and the same driver may change his or her driving
behavior over
time)~\cite{Ossen-interDriver06,tordeux2010adaptive,ossen2011heterogeneity},
this gives more control when investigating the models.

\begin{figure}
\centering
\includegraphics[width=0.9\textwidth]{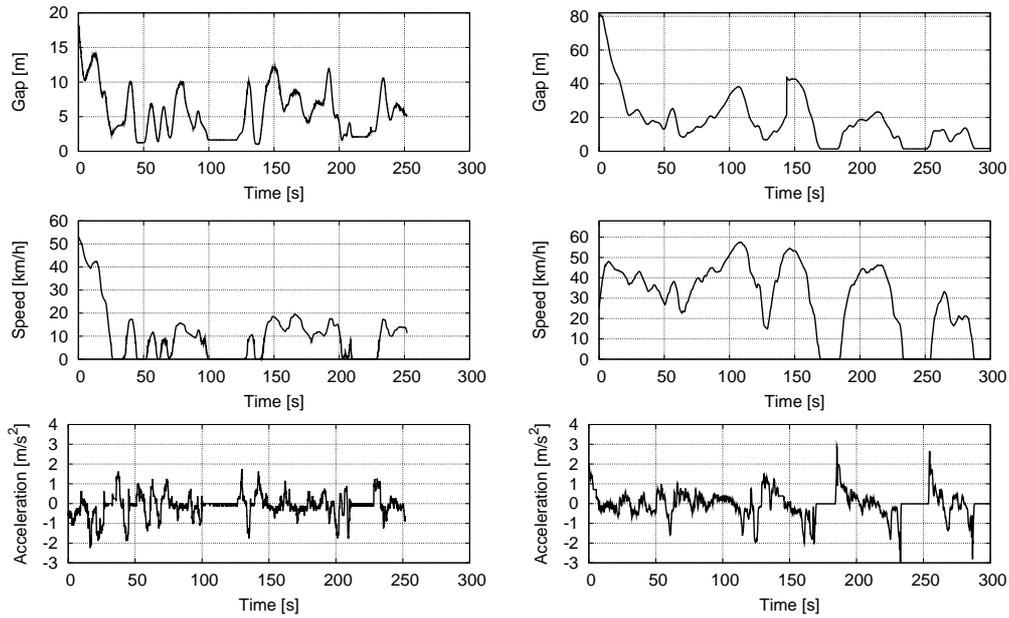}
\caption{\label{fig:Bosch-xFCD}Sets~1 and~3 of extended floating-car data of
a car driving in a German city during rush-hour conditions. In the set
labelled~3, 
the leader leaves the lane at $t\approx \unit[144]{s}$ resulting in a
discontinuity in the time series for the gap.
}
\end{figure}

We consider two categories of individual-driver data:
\emph{Extended floating-car data} (xFCD) come in the form of time series for
the  (arc-length) positions $x_i=x\sup{data}(t_i)$, 
speeds $v_i=v\sup{data}(t_i)$, and (bumper-to-bumper) gaps $s_i=s_i\sup{data}(t_i)$
at times $t_i=t_0+i \, \Delta t$. Here, $i$ denotes the time step and
$\Delta t$ is the sampling interval (inverse of the sampling rate). Since
$\abl{x}{t}=v$,
 positions and speed are not
independent and only one of them  can be
considered as the primary quantity and the other as derived quantity. In any case, the gap is measured
independently and directly by a range sensor. In this contribution, we use three
sets of xFCD which are captured by an instrumented car (sampling rate
10 Hz) driven 
through a German inner-city street by three different drivers
(cf. Fig.~\ref{fig:Bosch-xFCD}).
%
 Since
this vehicle was not equipped with high-precision (differential) GPS,
we consider the speed and gap measurements as primary data, i.e., we
derive the positions (or rather, displacements)
 of the follower and the leader from the speed of
the follower and the gap (between leader and follower) rather than vice versa.

\begin{figure}
\centering
\includegraphics[width=0.7\textwidth]{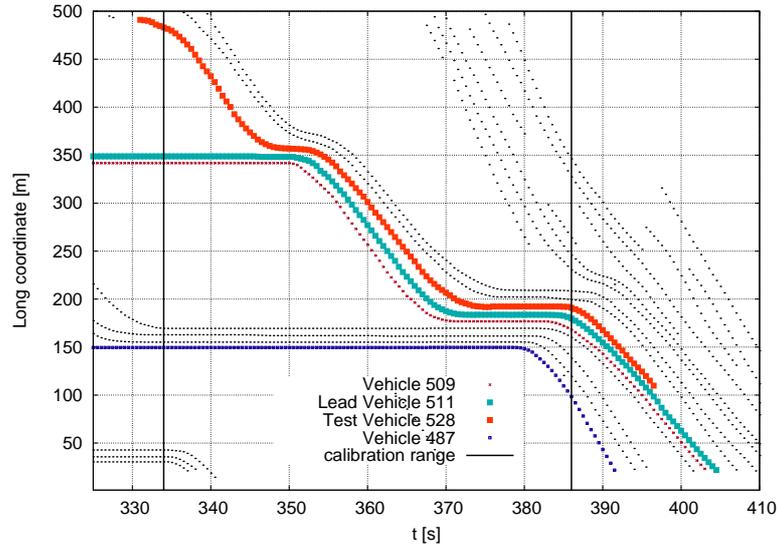}
\caption{\label{fig:NGSIM}Example of trajectories from the NGSIM project.
  Shown is a spatiotemporal region of the Lankershim set (every 5th data
  point is displayed) for
  direction~4 and lane~2. Vehicles~528 and 509 represent ``complete''
  data containing all traffic situations while Vehicle~487 
  includes minimum-gap and acceleration information, only. Stopping
  lines of traffic lights are located at approximatively 
  \unit[145]{m} and \unit[340]{m}. Vehicle~528
  with its leader~511 will be investigated in the calibration example later
  on. All other trajectories in this region are dotted.
}
\end{figure}

\emph{Trajectory data} provide the
locations (and lanes) of all vehicles in a given spatiotemporal
region, so it is straightforward to extract the (arc-length) positions $x_i$ and
$x_{li}$ of the considered vehicle and of its leader,
respectively. Several freeway and arterial data sets are made
available by the NGSIM project~\cite{NGSIM}.
It turned out that the data cannot be used directly since they contain
all sorts of inconsistencies and
errors~\cite{Thiemann-NGSIM-TRB08,punzo2011assessment}  which significantly influences
the calibration results, particularly for the less robust
models. Figure~\ref{fig:NGSIM} shows an extract of these data where no
obvious discrepancies are observed. 

The NGSIM data, and to a lesser extent the inner-city
xFCD data, contain negative speeds, negative gaps, unreasonable
values for accelerations, unreasonable frequency of sign changes for
accelerations, and sudden ``jumps''
of vehicles forwards, backwards, or to the side. Furthermore, both the xFCD and
NGSIM data contain positions, speeds, and accelerations, i.e., 
partially redundant information, and it is not always documented
how the dependent quantities are determined. Moreover,
they generally violate basic kinematic
constraints, particularly, that of
\emph{internal consistency},
\begin{equation}
\label{internalConsistency}
\abl{x}{t}=v(t), \quad \abl{v}{t}=\dot{v}(t),
\ee
and that of \emph{platoon consistency}~\cite{punzo2009estimation}
\be
\label{platoonConsistency}
s(0)=x_l(0)-x(0)-l_l, \quad
\abl{s}{t}=v_l(t)-v(t).
\ee
Here, $x(t)$ and $x_l(t)$ are the arc-length positions of the front
bumper of the considered and leading vehicles, respectively, $v(t)$
and $v_l(t)$ the corresponding speeds,
$s(t)$ denotes the bumper-to-bumper gap in between, and $l_l$ denotes
the length of the leading vehicle.
We focus on calibrating car-following models of the generic form
\be
\label{micGen}
\dot{v}(t) \equiv \abl{v}{t}=a\sub{mic}(s,v,v_l,\dot{v}_l)
\ee
returning as endogeneous variable (dependent or output variable) 
 the acceleration $\dot{v}(t)$ as a function  of the
gap and the speeds of the considered and leading vehicles. The
function $a\sub{mic}(.)$ specifies the car-following model in
question. Some models
such as the ACC model~\cite{kesting-acc-roysoc} also include the
leading acceleration  $\dot{v}_l$. If reaction times are modelled
explicitely, some or all of the above exogeneous variables
(also denoted as independent or input model variables) are taken at a delayed time.
Discrete-time models such as Gipps' model can be cast into this
form by formulating the acceleration in terms of finite speed
differences which these models provide, $\dot{v}(t)=[v(t+\Delta
t\sub{sim})-v(t)]/\Delta t\sub{sim}$ where $\Delta
t\sub{sim}$ denotes the model update time interval which is not
necessarily identical to the interval $\Delta t$ between two data points.
In order to obtain consistent training data sets for calibration,
i.e., time series of the exogeneous model variables $s(t)$, $v(t)$, and
$v_l(t)$ as well as the endogeneous acceleration,
we ignore all redundant (and inconsistent) information provided by the
data sets. Instead, we calculate it from the directly measured data
using the consistency criteria. 

As already mentioned, the directly measured xFCD quantities are the gaps $s_i$ (measured by the
range sensor) and the speed values $v_i$. After setting negative gap
and speed values equal to zero, we calculate the remaining variables
using Conditions~\refkl{internalConsistency}
and~\refkl{platoonConsistency} by the following  finite
differences:
\be
\label{xFCDcompletion}
v_{li}=v_i+\frac{s_{i+1}-s_{i-1}}{2\Delta t}, \quad
\dot{v}_i=\frac{v_{i+1}-v_{i-1}}{2\Delta t}, \quad
\dot{v}_{li}=\dot{v}_i+\frac{s_{i+1}-2s_i+s_{i-1}}{(\Delta t)^2}.
\ee
Discontinuities of the gap time series of xFCD are not necessarily
artifacts but may be the result of active lane changes (the driver of the instrumented
vehicle changes lanes), passive cut-out lane changes (the leader
changes to another lane as in Set 3 of Fig.~\ref{fig:Bosch-xFCD} at
$t\approx \unit[144]{s}$),  or passive cut-in changes
(a new vehicle changes into the gap before the instrumented vehicle). 
We notice that positions and vehicle lengths are not contained in the
set of exogeneous
variables of the considered class of car-following models. These quantities are neither
known for the investigated xFCD data nor needed for microscopic calibration.

The primary quantities of the NGSIM data are the positions,
particularly the positions $x_i$ and $x_{li}$ of a pair of vehicles
following each other at times $t_i=t_0+i\Delta t$. We calculate
the gap by
\be
\label{s_from_x}
s_i=x_{li}-x_i-l_i.
\ee
Notice that we require that the data contain the vehicle length of the
 leader and a definition whether $x$ denotes the position of the
 front bumper (for which the above relation is
 valid), or another well-specified position. Afterwards, we check for negative gaps
 or backwards moving vehicles and calculate the dependent quantities
 $v_i$, $v_{li}$, and $\dot{v}_i$ by the appropriate
 symmetric finite differences,
\be
\label{trajcompletion}
v_i=\frac{x_{i+1}-x_{i-1}}{2\Delta t}, \quad
v_{li}=\frac{x_{l,i+1}-x_{l, i-1}}{2\Delta t}, \quad
\dot{v}_i=\frac{x_{i+1}-2x_i+x_{i-1}}{(\Delta t)^2},
\ee
 thereby ensuring internal and platoon consistency.  

When testing the effects of different sampling rates or data
smoothing (Sec.~\ref{sec:smooth}), we resample the
primary data and/or 
apply kernel-based moving
averages (with Gaussian kernels of a certain width $w$) on them before 
calculating the remaining quantities with~\refkl{xFCDcompletion}
 or~\refkl{trajcompletion}, respctively.
In this way, we ensure
that the smoothed or re-sampled data are consistent as well.


Finally, in order to have more control over the traffic situations in
Sec.~\ref{sec:smooth}, we also use
generated data of vehicle platoons produced
 by microscopic simulations with a specified car-following model and a
 prescribed speed
profile for the leading vehicle (cf.
Fig.~\ref{fig:virtualxFCD} below). By virtue
of the simulation, these data are automatically consistent. 
Calibrating a model to data produced by the \emph{same} model is
the only methodologocal means to undertake investigations requiring
knowledge of the
\emph{true} parameters, e.g., when investigating the
bias introduced by smoothing or noise (Section~\ref{sec:smoothing}).

\section{\label{sec:calib}Approaches to Calibration}

We investigate two approaches: In the \emph{local approach}, the
endogenous model variable (speed change or acceleration) is compared
against the data, separately for each data point, by maximum-likelihood
techniques~\cite{hoogendoorn2010calibration}. In the
\emph{global approach}, we compare a complete data trajectory 
with a simulated trajectory and define an objective function in terms
of a sum of squared errors (SSE) of gaps or speeds.
Since we focus on model properties rather than on real-time traffic
state estimation, we will not investigate calibration by 
extended Kalman filters
where the parameters and the traffic state are estimated
simultaneously by maximum-likelihood
techniques~\cite{punzo2005part,Wang-RealTimeEstimation-2005,wang2008real}.

While the local approach is purely model based, the global approach
can be considered as simulation based~\cite{ciuffo2008comparison}: 
Each calculation of the objective function involves
a complete simulation run. 

\subsection{Local Maximum-Likelihood Calibration}

This approach assumes explicit noise of a given distribution, either
in the model (stochastic car-follownig models), or in the data, or in both. 
For each time instance $t_i$, the data contain the vector $\vec{x}_i$
of all exogeneous
variables needed for the model and the vector $\vec{y}_i$ of
all endogeneous variables.  This enables us to define the likelihood
function as the joint probability that the
model predicts \emph{all} data points given a certain parameter vector
$\vecbeta$:
\be
\label{def-Lgen}
L(\vecbeta)=\text{prob}\left(\hatvec{y}_1(\vecbeta)=\vec{y}_1, \ \ldots, \  
\hatvec{y}_n(\vecbeta)=\vec{y}_n\right).
\ee
If we assume
zero-mean Gaussian multivariate noise with fixed but unknown covariance matrix $\Sigma$ 
which is iid with respect to
different time steps (no serial correlation), the
log-likelihood $\tilde{L}(\vecbeta)=\ln L(\vecbeta)$
becomes~\cite{hoogendoorn2010calibration} 
\be
\label{logL-Gaussian}
\tilde{L}(\vecbeta,\Sigma)=\text{const.}
 - \frac{n}{2} \ln (\text{det} \, \Sigma)
 - \frac{1}{2} \sum\limits_{i=1}^n \vec{e}_i\tr(\vecbeta) \, \Sigma^{-1}\vec{e}_i(\vecbeta)
\ee
where
\be
\label{def-e}
\vec{e}_i(\vecbeta)=\hatvec{y}_i(\vecbeta)-\vec{y}_i
\ee
denotes the vector of deviations at time $t_i$. Estimating the covariance matrix by
\be
\label{ML-cov}
\hat{\Sigma}(\vecbeta)=\frac{1}{n}\sum\limits_{i=1}^n
 \vec{e}_i(\vecbeta) \, \vec{e}_i\tr(\vecbeta)
\ee
allows us to formulate the log-likelihood
\be
\label{logL-explicit}
L^*(\vecbeta)=\tilde{L}\left( \vecbeta,\hat{\Sigma}(\vecbeta) \right)
\ee
as a function of the parameter vector  $\vecbeta$ alone. This 
is the basis for parameter estimation:
\be
\label{MLgen}
\hatvecbeta=\text{arg} \ \max L^*(\vecbeta).
\ee

When calibrating models of the form~\refkl{micGen} on the motion of
single vehicles, 
there is only one endogeneous variable which is given by
the acceleration, i.e., $\hatvec{y}_i=a_{i}\sup{mic}$ and
$\vec{y}_i=\dot{v}_{i}$ for the predictions and observations, respectively.
Augmenting models of the form~\refkl{micGen}
by iid acceleration noise,
\be
\abl{v}{t}=a\sub{mic}(s,v,v_l; \vecbeta)+\epsilon, \quad
\epsilon \sim iid \; N(0, \sigma^2),
\ee
or assuming that the observed accelerations have this sort of noise,
we can derive from~\refkl{MLgen} the explicit calibration condition
\be
\label{MLacc}
\hatvecbeta=\text{arg} \ \min S\sub{ML}(\vecbeta), \quad
S\sub{ML}(\vecbeta)
=\sum\limits_{i=1}^n \left(\dot{v}_i-a_i\sup{mic}(\vecbeta)\right)^2\,,
\ee
where $a_i\sup{mic}(\vecbeta)=a\sub{mic}(s_i, v_i, v_{li},
\dot{v}_{li}; \vecbeta)$ and $S\sub{ML}(\vecbeta)$ is related (but not
identical) to $-\ln L(\vecbeta)$.
Since the statistical properties are specified explicitely,
the statistical properties of the parameter estimates are, in
principle, known:
Its covariance matrix is given in terms of the inverse Hessian of
$-\ln L(\vecbeta)$ which is provided by most nonlinear optimization packages.
However, in contrast to the iid assumption in deriving this matrix,
the deviations $e_i$ are serially correlated which 
can be seen by comparing the estimated and observed accelerations
of Fig.~\ref{fig:calLoc-IDM}. Consequently, the error
estimates are not valid. As we will show in Sec.~\ref{sec:smooth},
even the estimates of the parameters themselves
may be biased.

\begin{figure}
\centering
\includegraphics[width=0.9\textwidth]{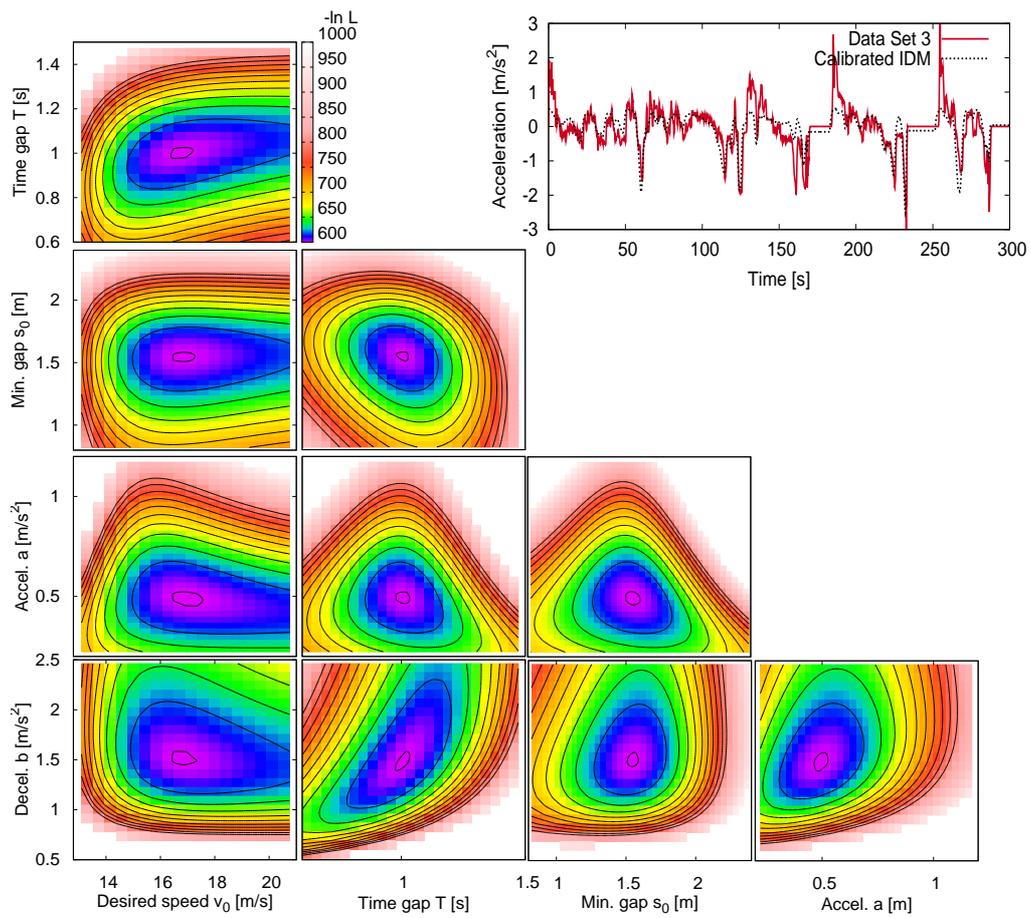}
\caption{\label{fig:calLoc-IDM}Local maximum-likelihood calibration of
  the IDM to the xFCD Set~3.
  Shown are the model accelerations for the calibrated parameters
shown in the left column of Table~\ref{tab:calib-IDM-Set3},
and contour plots of the objective function~\refkl{MLacc} for
two-dimensional sections of the five-dimensional parameter space
around the estimation. 
}
\end{figure}

\begin{table}
\begin{center}
\begin{tabular}{|l||c|c|c|} \hline
IDM parameter & local calibration
 & global calibration~\refkl{obj-s} 
 & global calibration~\refkl{obj-lns} \\ \hline
desired speed $v_0 [\unit[]{m/s}]$  & 16.9  & 16.1  & 16.1 \\
desired time gap $T [\unit[]{s}]$      & 1.02  & 1.27  & 1.20 \\
minimum gap $s_0 [\unit[]{m}]$    & 1.56  & 1.44  & 1.53 \\
maximum acceleration $a [\unit[]{m/s^2}]$  & 0.51  & 1.46  & 1.39 \\
comfortable deceleration $b [\unit[]{m/s^2}]$  & 1.47  & 0.63  & 0.65 \\ 
\hline
\end{tabular}
\end{center}
\caption{\label{tab:calib-IDM-Set3} calibration of the IDM to the
  xFCD Set~3 with respect to three objective functions.}
\end{table}

Figure~\ref{fig:calLoc-IDM} shows the result when calibrating the
five parameters of the intelligent-driver model (IDM) \cite{Opus}
to the xFCD Set~3 (cf. Fig.~\ref{fig:Bosch-xFCD}). We observe that the ``fitting
landscape'' of the objective function $S\sub{ML}(\vecbeta)$ is smooth
and contains a unique global 
minimum. This allows us to apply the efficient Levenberg-Marquardt
algorithm for solving
the nonlinear multivariate minimization problem~\refkl{MLacc}. We have 
used the open-source package \texttt{levmar}\cite{levmar}. A
calibration of a five-parameter model to the $\approx 3\,000$ xFC data
points  takes a few milliseconds. As result of the calibration, 
four of the five estimated IDM
parameters (desired speed $v_0$, desired time gap $T$, minimum gap
$s_0$, and comfortable deceleration $b$) assume
plausible values while the value of the acceleration parameter $a$ is
unrealistically low.  

\subsection{\label{sec:calGlob}Global Least-Squared Errors Calibration}

In the more commonly used global approach to trajectory calibration, 
we do not calibrate the model
directly by independently comparing the endogeneous model variables
with the observations, step by step. Rather, we use simulations of the
model to obtain predicted trajectories which, then, are compared with
the observed trajectories by formulating 
objective functions  in terms of sums of squared errors
 (SSE) of appropriate variables $\vec{y}$,
\be
\label{SSEsimple}
S(\vecbeta)=\sum\limits_{i=1}^n
\left(\hatvec{y}_i(\vecbeta)-\vec{y}_i\right)^2.
\ee
Similarly to the local ML approach, minimizing this function  reduces the calibration
problem to a multi-variate nonlinear optimization 
problem. In fact, \refkl{SSEsimple} includes the objective
function~\refkl{MLacc} by setting $\vec{y}_i=\dot{v}_i$. However, it
turns out that neither accelerations nor speeds are suited as basis
for the SSE of the simulation-based global approach since a
speed-based measure would be insensitive to parameters controlling the
gaps (e.g., $s_0$ and $T$ for the IDM), and acceleration-based
measures are additionally insensitive to the desired speed $v_0$. This
agrees also with the findings of Punzo and Simonelli~\cite{punzo2005analysis}. 
We therefore investigate absolute and relative measures of the gap
differences by defining the objective functions
\begin{eqnarray}
\label{obj-s}
S_s\sup{abs}(\vecbeta) &=& \sum\limits_{i=1}^n
\left(\hat{s}_i(\vecbeta)-s_i\right)^2, \\
\label{obj-lns}
S_s\sup{rel}(\vecbeta) &=& \sum\limits_{i=1}^n
\left(\ln \hat{s}_i(\vecbeta)-\ln s_i\right)^2
= \sum\limits_{i=1}^n
\left[\ln \left(\frac{\hat{s}_i(\vecbeta)}{s_i}\right)\right]^2
\end{eqnarray}
When applying~\refkl{obj-s}, the focus is on larger gaps corresponding
to acceleration or free-flow regimes (for the IDM, this relates to the
 parameters $v_0$ and $a$) while~\refkl{obj-lns} considers all
 regimes. In contrast
 to~\cite{Kesting-Calibration-TGF07,Arne-DissOnline}, we use as measure of the
 relative deviation
 the logarithm $e_i=\ln(\hat{s}_i/s_i)$ rather than the ratio 
$(\hat{s}_i-s_i)/s_i$ . While both measures are equivalent to first
 order,
\be
e_i= \ln\left(\frac{\hat{s}_i}{s_i}\right)=
\frac{\hat{s}_i-s_i}{s_i} + {\cal O}
\left(\frac{\hat{s}_i-s_i}{s_i}\right)^2,
\ee
the objective function~\refkl{obj-lns} is less sensitive to outliers
and measuring errors. 

In order to obtain the simulation-based estimates $\hat{s}_i$, 
we initialize the speed of the simulated
 vehicle and its initial gap to the leading vehicle by the data and let 
the vehicle follow the fixed observed speed profile 
$v_l(t)$ of its leader as obtained by~\refkl{xFCDcompletion}:
\be
\label{calibMicGlobal}
\abl{\hat{v}}{t}=a\sub{mic}(\hat{s},\hat{v},v_l; \vecbeta), \quad
\hat{s}(t_0)=s(t_0), \ \hat{v}(t_0)=v(t_0).
\ee
We treat discontinuities caused by active or passive lane changes (as
in the xFCD Set~3 at $t\approx\unit[144]{s}$) in the same way as a
range sensor of an adaptive cruise control system
detecting a new target would do, i.e., by introducing a data-driven discontinuity  of
the simulated  gap:
\be
\label{newTarget}
\hat{s}(t^+)=\hat{s}(t^-) + s(t^+)-s(t^-)
\ee 
where $t^-$ and $t^+$ denote the times immediately before and after
the detection of the new target, respectively. To determine whether
such a discontinuity has occurred, we compare the ballistic extrapolation 
\be
\hat{s}\sub{ball}(t_i+\Delta t; \Delta a)=s_i+(v_{li}-v_i)\Delta t + \frac{1}{2}\Delta a
(\Delta t)^2
\ee
with the observation $s_{i+1}$.
The event ``a new target
is detected'' is triggered if $\hat{s}\sub{ball}(t_i+\Delta t; \Delta a)=s_{i+1}$ is not
true for any
acceleration difference in the range $|\Delta a|<\unit[20]{m/s^2}$. 

Instead of~\refkl{newTarget}, we also tested a ``hard reset'' according to
\be
\label{newTargetReset}
\hat{s}(t^+)=s(t^+), \quad \hat{v}(t^+)=v(t^+)
\ee 
which turned out to produce little 
differences. The reset has the advantage that multiple trajectories
can be calibrated simultaneously by concatenating the
corresponding data and evaluating the objective function by 
running a single car-following simulation which
is controlled by the resulting compound xFCD via~\refkl{calibMicGlobal}
and~\refkl{newTargetReset}.
\begin{figure}
\centering
\includegraphics[width=0.9\textwidth]{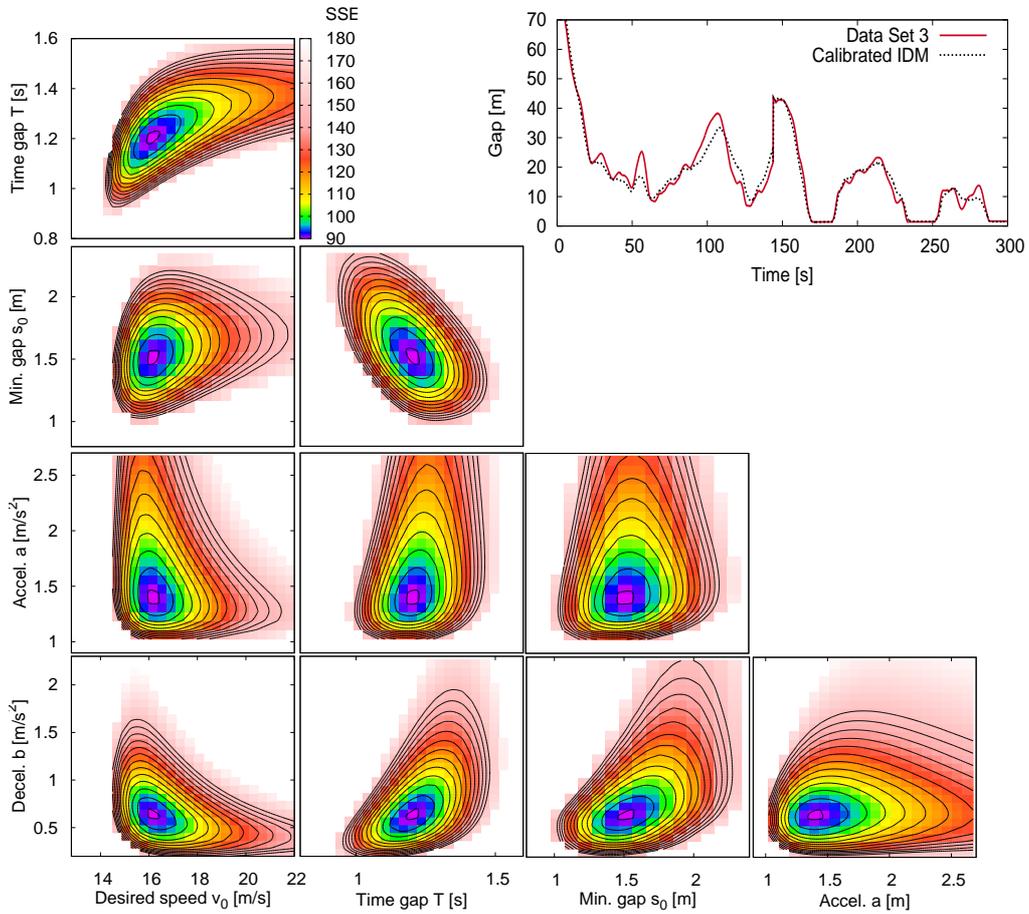}
\caption{\label{fig:calGlob-lns-IDM}Global calibration of the IDM to the
  xFCD Set~3 with respect to the logarithms of the gaps.
  Shown are the observed and simulated gaps for the calibrated parameters
(Table~\ref{tab:calib-IDM-Set3}, right column) 
 and contour plots of the objective function~\refkl{obj-lns} for
two-dimensional sections of the five-dimensional parameter space
around the estimation. 
}
\end{figure}

\begin{figure}
\centering
\includegraphics[width=0.8\textwidth]{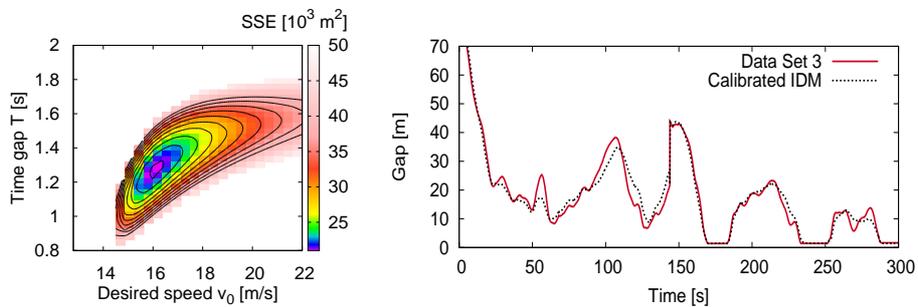}
\caption{\label{fig:calGlob-s-IDM}Global calibration of the IDM to the
  xFCD Set~3 with respect to the gaps.
}
\end{figure}

\begin{figure}
\centering
\includegraphics[width=1.0\textwidth]{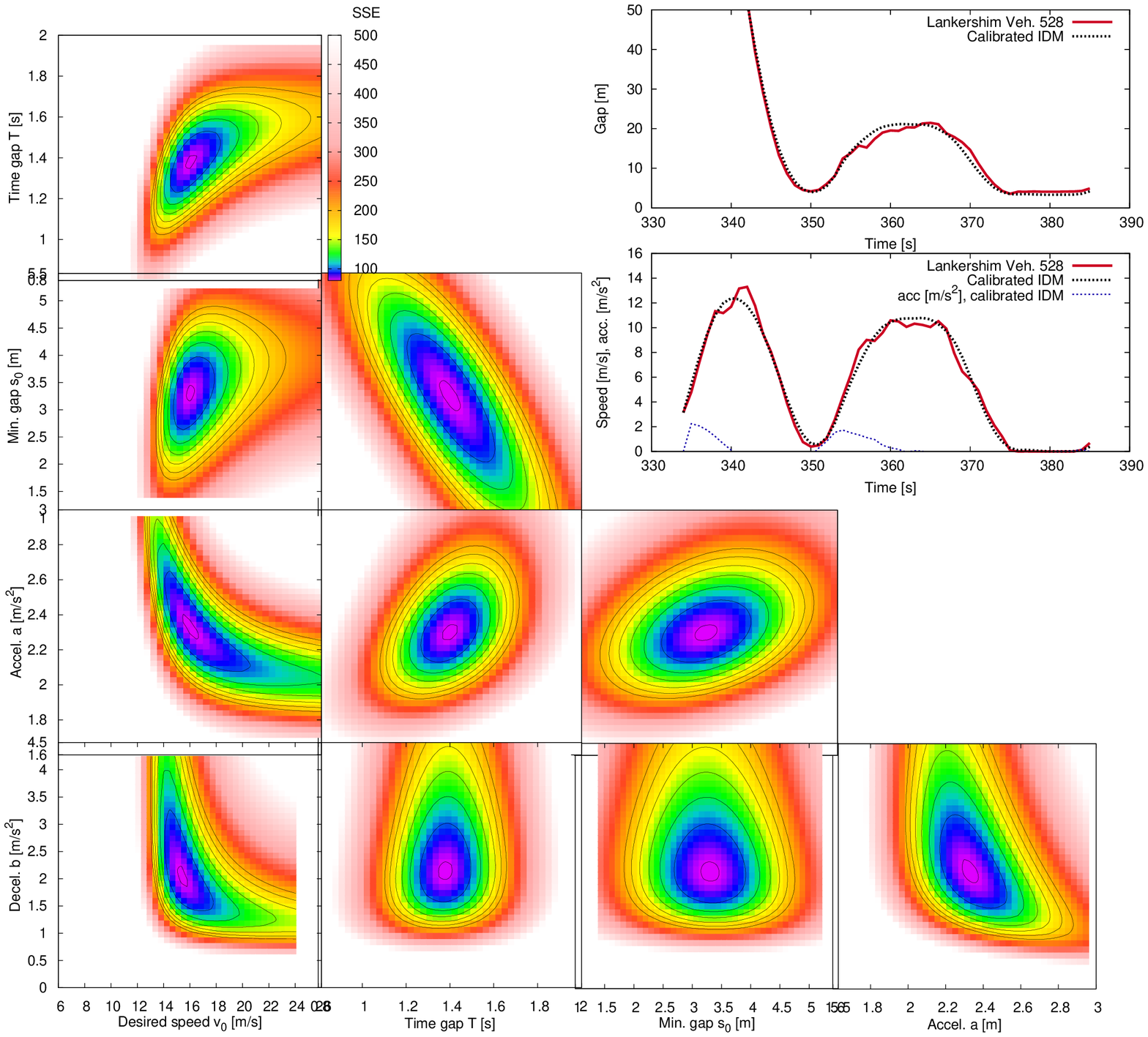}
\caption{\label{fig:NGSIM-calib}Global calibration of Vehicle~528 of the
  NGSIM Lankershim data set (for the trajectories, see
  Fig.~\ref{fig:NGSIM}) with respect to the gap deviations in the time interval
  [\unit[334]{s},\unit[386]{s}]. Shown is the fitting quality with
  respect to the gaps and speeds and two-dimensional subsets of the objective function
  around the calibrated parameters. Obviously, the objective function is smooth
  and unimodal.
}
\end{figure}

Finally, we need to treat the case that the range of suitable or prescribed 
simulation update time intervals $\Delta t\sub{sim}$ is incompatible
with the sampling interval $\Delta t$. This applies to most of the time-discrete
car-following models (e.g., Gipps' model~\cite{Gipps81} or Newell's
car-following model~\cite{newell-carFollowing2002}) which have  their update
time step fixed by a model parameter (the desired time gap or the
reaction time). On the
other hand, some
time-continuous models such as the optimal-velocity model (OVM)~\cite{Bando1}
often require time steps below~\unit[0.1]{s} for a stable
simulation~\cite{TreiberKesting-Book}. In this case, we run the
simulation with the appropriate time step $\Delta t\sub{sim}$
resulting in the times $t_j\sup{sim}=t_0+j\Delta t\sub{sim}$. Since
$t_j\sup{sim}$ generally is not commensurable with the data times
$t_i$, we evaluate the simulation control~\refkl{calibMicGlobal}
and~\refkl{newTarget} (or~\refkl{calibMicGlobal} and ~\refkl{newTargetReset}) at the times
$t_j\sup{sim}$ by a piecewise linear
interpolation of the data.
Conversely, we determine the deviations $e_i$ between the simulation
and the data at the data time instants $t_i$
by a piecewise linear interpolation of the simulation results.

The Figures~\ref{fig:calGlob-lns-IDM} and~\ref{fig:calGlob-s-IDM} show
global calibrations of the IDM to xFCD Set~3 with respect to the
relative and absolute gap
differences. In the calibrations, the full period (\unit[300]{s})
available in the data has been used for calibration.
 As for the local calibration, the ``fitting
landscapes'' of the objective functions~\refkl{obj-lns}
and~\refkl{obj-s} are smooth and unimodal, so we can apply the
Levenberg-Marquardt algorithm for solving the optimization problem. 
Generally, the local approach is faster but also the global approach
takes only about \unit[20]{ms} to estimate a five-parameter model to
3\,000 data points. The differences of
the estimated parameters between the two global objective functions
turn out to be marginal. However, the estimates show
significantly differences with respect to the local ML calibration. 
Particularly, the estimates of the acceleration parameters
$a$ and $b$ have swapped their magnitude with $b$ now being lower than
expected while the maximum acceleration $a$ now concurs with
expectations. The low value of $b$ can possibly be explained by
observing that the IDM parameter $b$ denotes the \emph{desired}
deceleration while the actual IDM deceleration can be higher if
required by the situation. Consequently, the desired deceleration is
unobservable if the data contain only car-following situations rather
than ``approaching situations'', i.e., transitions from free traffic
flow to car-following. 
In fact, data Set~3 shows only one approaching situation in the first
\unit[20]{s}. In this episode, the averaged observed acceleration of
$\approx\unit[-0.5]{m/s^2}$ - $\unit[-0.5]{m/s^2}$ is comparable to the parameter $b$. 
This is also confirmed by calibrating Gipps's model. The estimate for
its deceleration parameter $b=\unit[1.01]{m/s^2}$ is lower than
expected as well, although the discrepancy is less pronounced.

A closer inspection of the local calibration shows that
it is sensitive to data noise which might be a cause of its unrealistically
low acceleration estimates: Uncertainties of positional and speed
measurements magnify in the acceleratione estimates due to
differentiation, so obviously, an objective based on the squares of
acceleration differences is not robust. Moreover, even an
acceleration parameter $a=0$ (the vehicle neither accelerates nor
decelerates and the leader is completely ignored) 
will lead to a reasonable fit while, in the global
simulation, such a  vehicle could crash or stand all the time (depending
on the initialization).

Figure~\ref{fig:NGSIM-calib} shows a global calibration of the IDM to
a single NGSIM
trajectory (Vehicle~528 of Fig.~\ref{fig:NGSIM}) 
of the Lankershim data set (representing a city arterial with
traffic-light controlled intersections)  with
respect to the objective~\refkl{obj-s}. This trajectory is complete in
the sense discussed in Section~\ref{sec:dataCompleteness}, i.e., it
contains all relevant traffic situations: freely accelerating
($t<\unit[340]{s}$), cruising at about the desired speed
($\unit[340]{s}\le t<\unit[342]{s}$), approaching a standing
vehicle from a large distance ($\unit[342]{s}\le t<\unit[350]{s}$), accelerating behind a
leader ($\unit[350]{s}\le t<\unit[360]{s}$), following a leader in
near steady state ($\unit[360]{s}\le t<\unit[365]{s}$),
decelerating behind a leader ($\unit[365]{s}\le t<\unit[375]{s}$), and
standing ($t>\unit[375]{s}$). The plots of the objective
function show that all IDM parameters are relevant. Moreover, although
we did not eliminate the known data imperfections and fluctuations in
the NGSIM data, there is strong evidence that the global objective
function is both smooth and unimodal.  

Further
calibrations show that not all parameters can be estimated if a
relevant situation is missing. For example, when restricting the
trajectory to be calibrated to the interval between \unit[350]{s} and
\unit[386]{s}, only lower limits of the desired speed $v_0$ and acceleration
parameter $a$ can be estimated. Furthermore, it turned out that even the global
calibration is sensitive to behavioural changes or exogeneous factors
not considered in the model. Specifically, for $t<\unit[334]{s}$,
Vehicle~528 only accelerates slowly although there is a free road
ahead and the speed is low (the traffic light just turned
green). Moreover, for $t>\unit[386]{s}$, the
gap \emph{decreases} while the speed is increasing. While this is
probably due to an anticipative action in preparation for 
an active lane change at $t\approx \unit[397]{s}$, the calibration of
this time interval would result in a \emph{negative} value of the time gap parameter
$T$ (or to significantly larger errors when calibrating the complete
trajectory) when ignoring this behavioural change.

After a series of further calibrations with various models and data (some of
which will be discussed below) we conclude that the global calibration
is conceptionally more reliable than the local
calibration. This is mainly due to the serial correlations of speed
and gaps which the simulation reproduces automatically.
Furthermore, constructing the objective function based on the absolute
or relative gap  is generally more robust, reliable,  and has a higher 
capability to differentiate between models, as measures based on speed
or accelerations\cite{punzo2005analysis}.

\section{\label{sec:smooth}Influence of the Data Sampling Rate and Smoothing}

Are data sampling intervals of $\Delta t=\unit[0.1]{s}$ really necessary in the
light of the obvious serial correlations of the dynamic variables? Does acceleration noise
(which is necessarily introduced
 by taking the numerical time derivative of the xFC speed data, or
 taking the second numerical
 derivative of the positional data of trajectory data) impair the
 result, i.e., is data smoothing necessary? If so, what is the optimal
 smoothing width? 
Our methods of data preparation (Sec.~\ref{sec:data}) and 
simulation procedures (Sec.~\ref{sec:calGlob}) allow us to
answer these question by a systematic sensitivity analysis of
the effects of different sampling intervals and data smoothing.

\begin{table}
\begin{center}
\begin{tabular}{|l||c|c|c|c|c||c|c|c|c|c|} \hline
 & \multicolumn{5}{|c||}{local calibration w.r.t. $S\sub{ML}(\vecbeta)$}
 & \multicolumn{5}{|c|}{global calibration w.r.t. $S_s\sup{rel}(\vecbeta)$}
 \\ \hline
\parbox{6em}{\vspace{0.3ex}

sampling\\[-0.3ex] interval $\Delta t$\\[-1.5ex]}
 &\unit[0.1]{s}&\unit[0.5]{s}&\unit[1.0]{s}&\unit[2.0]{s}&\unit[5.0]{s}
 &\unit[0.1]{s}&\unit[0.5]{s}&\unit[1.0]{s}&\unit[2.0]{s}&\unit[5.0]{s}
 \\ \hline
$v_0 [\unit[]{m/s}]$ & 16.9 & 16.8  & 16.8 & 16.6 & 16.1 & 16.1 & 16.2 & 16.3 & 15.8 & 14.8 \\
$T [\unit[]{s}]$     & 1.02 & 1.01  & 1.00 & 1.02 & 0.90 & 1.20 & 1.21 & 1.22 & 1.12 & 0.87 \\
$s_0 [\unit[]{m}]$   & 1.55 & 1.56  & 1.57 & 1.58 & 1.72 & 1.53 & 1.54 & 1.58 & 2.05 & 3.12 \\
$a [\unit[]{m/s^2}]$ & 0.52 & 0.51  & 0.49 & 0.43 & 0.31 & 1.39 & 1.38 & 1.37 & 1.35 & 1.24 \\
$b [\unit[]{m/s^2}]$ & 1.47 & 1.47  & 1.52 & 1.90 & 2.77 & 0.65 & 0.65 & 0.66 & 0.76 & 0.28 \\ 
\hline
error [m/s$^2$ or \%] & 0.45 & 0.42 & 0.40 & 0.38 & 0.36 & 17.4 & 17.2 & 17.7 & 19.9 & 32.2 \\
\hline
\end{tabular}
\end{center}
\caption{\label{tab:calib-IDM-sampling} Local and global calibration of the IDM to the
  xFCD Set~3 with different sampling intervals. Also shown is the fit
  quality in terms of the rms absolute acceleration and relative gap errors
for the local and global calibrations, respectively.}
\end{table}

\subsection{\label{sec:sampling}Sampling intervals}
%
We investigate the  effect of different
sampling intervals by eliminating the corresponding data points of the original
xFC data. For example, for a sampling width $\Delta
t=\unit[1]{s}$, only \unit[10]{\%} of the original data points are
retained, on which the calibration is performed.
Table~\ref{tab:calib-IDM-sampling} shows the result for a local and a
global IDM calibration. Remarkably, the results differ
only insignificantly with respect to a calibration on the original as
long as the sampling interval does not exceed one second.
 In order to assess whether the
statistical properties of the parameter estimates change with $\Delta
t$, we also look at the shape of the
fitting landscape. The top left contour plot of 
Fig.~\ref{fig:samplingSmoothing} displays a
representative section of the fitting landscape in the $v_0$ - $T$
plane for
 $\Delta t=\unit[0.1]{s}$. Comparing it with the corresponding plot of
Fig.~\ref{fig:calGlob-lns-IDM} shows that, in relative terms, the
fitting landscape does not change significantly when going from
$\Delta t=\unit[0.1]{s}$ to \unit[1]{s}. In fact, all correlation
coefficients of the parameter estimates change by less than
\unit[2]{\%}.

 Since the absolute values of the objective functions are essentially
 proportional to the number of data points, common statistical or
 optimization software packages will indicate that the error of the
 parameter estimates is proportional to the inverse square root of the
 number of data points. However, this is only true for iid errors
 which does not apply here. We have investigated the errors directly
 by calibrating the IDM to differently re-sampled data sets. For
 example, for $\Delta t=\unit[1]{s}$, we can take the data lines for
 $t_i=\unit[1]{s}$, \unit[2]{s}, ... but also the data corresponding
 to \unit[0.1]{s}, \unit[1.1]{s}, and so on. The resulting
 fluctuations of the parameter estimates for the different sets
 confirms that the error depends only insignificantly on the sampling
 interval  as long as
 $\Delta t \le \unit[1]{s}$. We conclude that a sampling interval of
 \unit[1]{s} is sufficient for calibrating car-following models to xFC
 or trajectory data. 

\begin{table}
\begin{center}
\begin{tabular}{|l||c|c|c|c||c|c|c|c|} \hline
 & \multicolumn{4}{|c||}{local calibration w.r.t. $S\sub{ML}(\vecbeta)$}
 & \multicolumn{4}{|c|}{global calibration w.r.t. $S_s\sup{rel}(\vecbeta)$}
 \\ \hline
smoothing kernel width $w$
 &\unit[0.5]{s}&\unit[1.0]{s}&\unit[2.0]{s}&\unit[5.0]{s}
 &\unit[0.5]{s}&\unit[1.0]{s}&\unit[2.0]{s}&\unit[5.0]{s}
 \\ \hline
$v_0 [\unit[]{m/s}]$ & 16.9  & 17.1 & 17.5 & 18.3 & 16.2 & 16.2 & 16.2 & 15.9 \\
$T [\unit[]{s}]$     & 1.02  & 1.03 & 1.05 & 1.08 & 1.21 & 1.22 & 1.22 & 1.24 \\
$s_0 [\unit[]{m}]$   & 1.56  & 1.57 & 1.58 & 1.69 & 1.54 & 1.54 & 1.57 & 1.67 \\
$a [\unit[]{m/s^2}]$ & 0.51  & 0.51 & 0.48 & 0.40 & 1.37 & 1.32 & 1.21 & 1.00 \\
$b [\unit[]{m/s^2}]$ & 1.41  & 1.29 & 1.03 & 0.62 & 0.69 & 0.69 & 0.68 & 0.58 \\ 
\hline
error [$m/s^2$ or \%] & 0.41 & 0.38 & 0.33 & 0.25 & 17.1 & 16.6 & 16.1 & 14.9 \\
\hline
\end{tabular}
\end{center}
\caption{\label{tab:calib-IDM-smooth} IDM calibrations to the
  original xFCD Set~3  with Gaussian
  kernel-based smoothing of different widths $w$.}
\end{table}

\begin{figure}
\fig{0.8\textwidth}{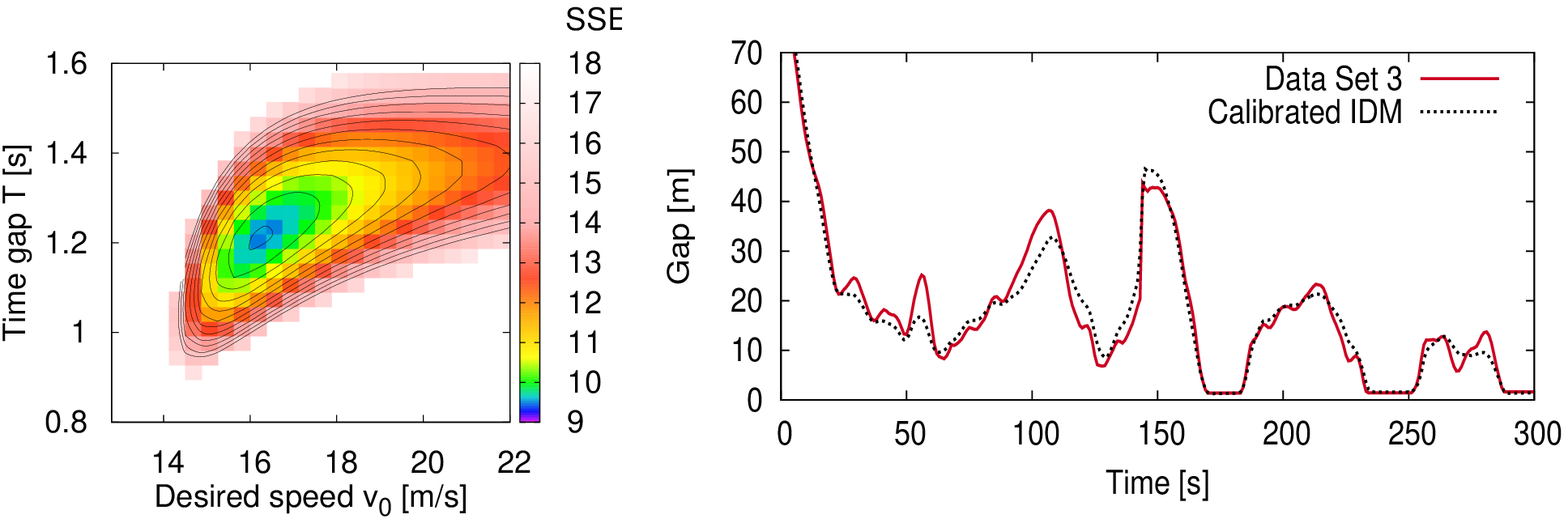}
\vspace{-5em}

\fig{0.8\textwidth}{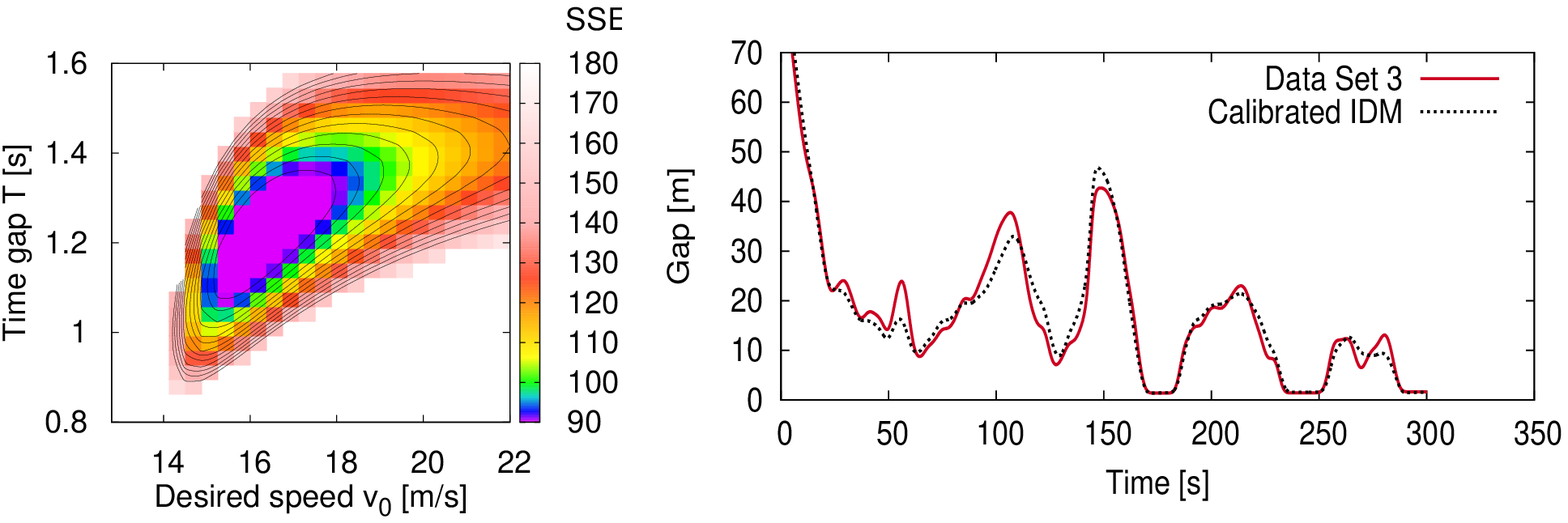}
\vspace{-2em}

\caption{\label{fig:samplingSmoothing}Sensitivity analysis of
the global calibration of the IDM
(objective function~\refkl{obj-lns}) to the xFCD Set~3 with respect to
data properties. Top: Sampling
interval $\Delta t=\unit[1]{s}$ instead of \unit[0.1]{s}. Bottom:
Gaussian kernel-based smoothing of the original set with a width
$w\unit[2]{s}$.
}
\end{figure}

\begin{figure}
\fig{0.8\textwidth}{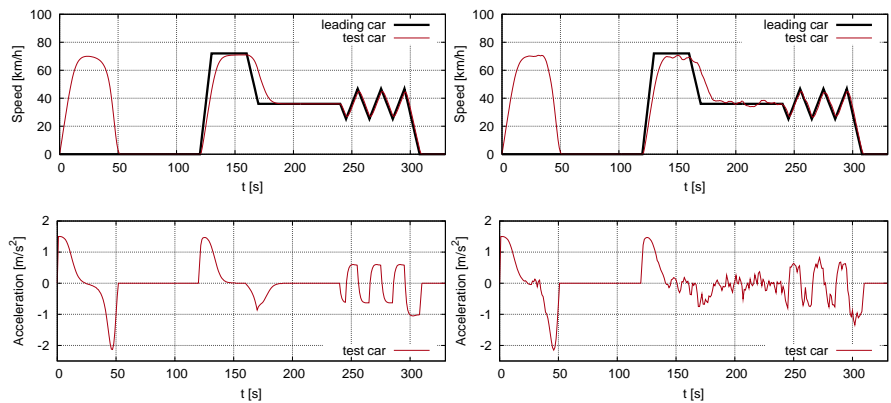}

\caption{\label{fig:virtualxFCD}virtual xFCD generated by an IDM vehicle
  ($v_0=\unit[20]{m/s}$, $T=\unit[1]{s}$, 
$s_0=\unit[2]{m}$, $a=\unit[1.5]{m/s^2}$, $b=\unit[2]{m/s^2}$) without
  noise (left), and with serially correlated noise induced -- via a model
  of human drivers~\cite{HDM} -- by estimation
  errors of the inverse time-to-collision $r=\Delta v/s$ (standard
  deviation~\unit[10]{\%}, 
  correlation time \unit[10]{s}) (right). The speed profile of the leading vehicle
 (thick solid line) and the initial gap and speed of the follower are prescribed. 
}
\end{figure}

\subsection{\label{sec:smoothing}Effects of Data Smoothing}

For the sensitivity analysis with respect to data smoothing, we start
with the original data (sampling interval~\unit[0.1]{s}), apply
Gaussian kernel-based averaging of variable smoothing width $w$ 
on the primary quantities
(gap and speed), and calculate the 
rest of the variables
by Eq.~\refkl{xFCDcompletion}. Table~\ref{tab:calib-IDM-sampling} and
Fig.~\ref{fig:samplingSmoothing} show that, up to a smoothing width
$w=\unit[1]{s}$, the estimates obtained by global calibration 
and the fitness landscape of the
objective functions (including correlations and covariances) remains
essentially unchanged. As expected, the fit quality increases
 with $w$ (while it deteriorates with the sampling interval for intervals above
 \unit[0.5]{s}).

As a further test to confirm this result, and also to test for a bias
of the parameter estimate caused by noise, we calibrate the IDM on
virtual xFCD generated by the same model with additional serially
correlated acceleration noise (Figure~\ref{fig:virtualxFCD}) and repeat the
calibration for several realizations of the noisy trajectory. We find no
significant bias 
with the exception of the acceleration parameter $a$ for local
calibration, which is systematically estimated too low. The observed
variations of the estimates, however, are higher than 
indicated by the optimization software confirming the influence of
serial correlations. Finally, 
data smoothing did not improve the result significantly.

We conclude that, in spite of possible 
artifacts due to numerical differentiation, no data smoothing is
necessary when performing a global calibration on xFC data. 
While first evidence indicates that this is also true for the NGSIM
data generally containing larger errors
(cf. Section~\ref{sec:calGlob}), a systematic investigation 
 remains to be done.

\section{\label{sec:complete}Investigations on the Parameter Space}

\begin{figure}
\fig{0.8\textwidth}{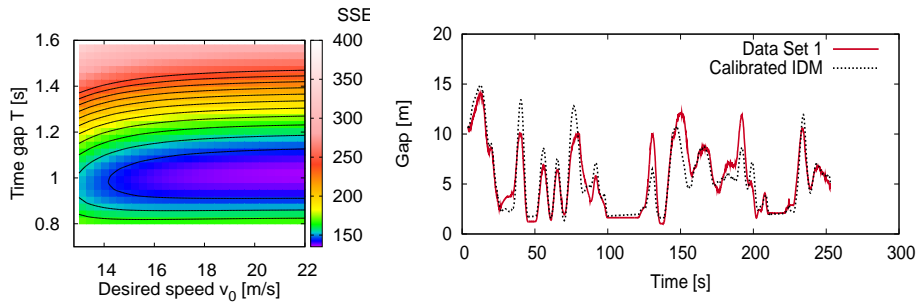}
\caption{\label{fig:Boschdata1}Global calibration of the IDM to xFCD Set~1
  with respect to the relative gaps.
}
\end{figure}

\subsection{\label{sec:dataCompleteness}Data completeness and parameter orthogonality}

When calibrating the IDM to the xFCD Set~3 (Fig~\ref{fig:Bosch-xFCD}),
we always obtain smooth objective functions with a unique global
minimum. This also is true for many other time-continuous models and
some time-discrete models. However, Fig.~\ref{fig:Boschdata1} shows
that things are different for xFCD
Set~1. The contour plot  of the objective landscape (left part of this
figure) shows that the minimum is indifferent with respect to the
desired speed. The same is true for the deceleration parameter $b$. In
fact, the unconstrained minimization yields $v_0\approx
\unit[1000]{km/h}$ (i.e., de facto, unbounded) and
$b=\unit[6.9]{m/s}$ which is unrealistically high as well. Qualitatively, we
obtain the same results for other models containing the desired speed
and/or the approaching deceleration as (combinations of) model
parameters as in Gipps' model. The reason lies
in the data set containing neither free-flow nor approaching
situations but only congested and stopped traffic.

In such a situation, it is favorable to use a model which is
\emph{parameter orthogonal} meaning that different identifiable driving situations
(such as freely accelerating, cruising, approaching, following or
standing) are represented by different parameters, ideally one for
each situation. If certain situations are not included in the
data, the corresponding parameters are unobservable and must not be
calibrated. Instead, we fix them to values which lie in the
corresponding indifferent regimes. The contour plot of Fig.~\ref{fig:Boschdata1}
indicates, that the desired speed lies in the indifferent regime if
$v_0>20$. Parameter orthogonal model allows to identify such regimes, a
priory, by the data. For the IDM, for example, it is safe to assume 
\be
\label{indifferentRegimes}
v_0=2  \max\limits_i\left(v_i \right), \quad
b = -2  \max\limits_i\left(-\dot{v}_i\right)
\ee
if the data contain no free-flow and no approaching situations, respectively. 
Calibrating the remaining IDM parameters with respect to relative gaps
on xFCD Set~1, we obtain 
$T=\unit[1.02]{s}$, $s_0=\unit[2.26]{m}$, and $a=\unit[1.22]{m/s^2}$ 
while the unconstrained five-dimensional calibration yielded 
$T=\unit[1.03]{s}$, $s_0=\unit[2.25]{m}$, and
$a=\unit[1.22]{m/s^2}$. We conclude that the values of the remaining
parameters and the fit quality (the minimized
SSE) do not change when restricting the
calibration. Additionally, the restriction reduces the indicative estimation
errors by about \unit[10]{\%}. 

Notice that, since the objective functions deviate strongly from their
local parabolic shape (see, e.g., Fig.\ref{fig:calGlob-lns-IDM}), 
parameter orthogonality does not necessarily mean that the
correlation matrix of the indicative estimation errors (or,
equivalently, the inverse Hessian of the objective function at the
estimated values) is diagonal. Nevertheless, low correlations not
exceeding absolute values of 0.5 indicate a good
model quality. For the IDM, the correlation between $v_0$ and $T$ is
often greater than 0.5 since the effects of increased desired speed
are partially compensated for by an increased following time
gap. However, all other cross correlations are generally below
0.5.

In order to apply~\refkl{indifferentRegimes}, we need a systematic
data-based procedure to distinguish the different regimes. We propose
five regimes and following discrimination criteria:
\bi
\item Cruising if the speed and the time gap are above data-driven
  limits $v_c$ and $T_c$, respectively, and the absolute acceleration
  is below a limit $a_c$,
\item free acceleration if the time gap is above $T_c$, the acceleration
  above $a_c$, and the conditions for cruising are not fulfilled,
\item approaching if $v>v_l$, the time gap $s/v>T_c$, and the
  kinematic deceleration $(v-v_l)^2/(2s)>a_c$,
\item essentially standing if the gap $s<s_c$,
\item following if $T<T_c$ and none of the above conditions applies,
\item and inconsistent behavior, otherwise.
\ei
%
\begin{table}
\begin{center}
\begin{tabular}{|l||c|c|} \hline
Regime            & xFCD Set~1   & xFCD Set~3    \\ \hline
cruising          & \unit[ 0]{\%}  & \unit[ 5]{\%}   \\
free acceleration & \unit[16]{\%}  & \unit[ 8]{\%}   \\
approaching       & \unit[ 2]{\%}  & \unit[ 3]{\%}   \\
standing          & \unit[ 4]{\%}  & \unit[15]{\%}   \\
following         & \unit[72]{\%}  & \unit[65]{\%}   \\
inconsistent      & \unit[ 3]{\%}  & \unit[ 4]{\%}   \\
\hline
\end{tabular}
\end{center}
\caption{\label{tab:regimes} driving regimes contained in the xFCD sets.}
\end{table}
%
Table~\ref{tab:regimes} shows the regime distribution for the two
considered xFC data sets. As expected, Set~1 does not contain cruising episodes and
describes approaching situations only \unit[2]{\%} of the time.

\subsection{Model completeness}
%
Not only data can be incomplete if they lack characteristic traffic
situations but also models if they are not able to describe all
situations. For example, car-following models in the stricter sense
cannot describe free traffic flow. An example of such an incomplete model
is the full velocity-difference model (FVDM)~\cite{Jiang-vDiff01}. This
model augments the optimal-velocity model~\cite{Bando1} by a linear
speed sensitivity term. It can be formulated as
\be
\label{VDIFF}
\abl{v}{t}=a\left(\frac{v\sub{opt}(s)-v}{v_0}\right) + \gamma (v_l-v).
\ee
Assuming an optimal-velocity function
\be
\label{OVF}
v\sub{opt}(s)=\max\left[0, \min \left(v_0,
    \frac{s-s_0}{T}\right)\right],
\ee
its static model  parameters $v_0$, $T$, and $s_0$ and the maximum
acceleration $a$ have the same meaning as the corresponding
IDM parameters. The IDM model parameter $b$ is replaced by the speed-difference
sensitivity $\gamma$.

The incompleteness is caused by the last term describing a linear sensitivity
to speed differences. Since this term does not depend on the gap, it
describes interactions of an infinite range, i.e., in contrast to the
OVM, the FVDM cannot describe free traffic flow~\cite{TreiberKesting-Book}.
%
\begin{table}
\begin{center}
\begin{tabular}{|l||c|c||c|c|} \hline
parameter & OVM, Set~1 & OVM, Set~3 & FVDM, Set~1 & FVDM, Set~3\\ \hline
desired speed $v_0 [\unit[]{m/s}]$        & 170  & 14.0  & 12.8& 14.1 \\
desired time gap $T [\unit[]{s}]$            & 1.09  & 1.38  & 1.08 & 1.44 \\
minimum gap $s_0 [\unit[]{m}]$          & 2.75  & 2.86  & 2.76 & 1.52 \\
maximum acceleration $a [\unit[]{m/s^2}]$        & 182  & 16.4  & 12.0 & 9.03 \\
relative speed sensitivity $\gamma [\unit[]{s^{-1}}]$   & -    & -    & 0.08 & 0.65 \\ 
\hline
\end{tabular}
\end{center}
\caption{\label{tab:OVM-VDIFF} Global calibration of the
  optimal-velocity and full-velocity-difference models~\refkl{VDIFF},
  \refkl{OVF} to the xFCD sets with respect to the relative gaps.}
\end{table}
%
Table~\ref{tab:OVM-VDIFF} shows that, as a consequence, the calibrated
FVDM essentially reverts to the OVM if such situations arise. As a further
consequence, the calibrated FVDM is nearly as unstable as the OVM 
and the acceleration parameter assumes unrealistically
high values in order to compensate for the increased instability by a
higher agility. This means, the FVDM is not only incomplete but
also not parameter orthogonal. Finally, we notice that, due to
the extreme values of the
calibrated OVM acceleration parameter, the OVM vehicles
follow their leader nearly statically at the optimal gap
$s\sub{opt}(v_l)=s_0+v_lT$ (the inverse of the optimal velocity
function) if the speed is below the desired speed.

\section{\label{sec:concl}Conclusions}

Most other publications on calibrating car-following models focus on
the fit quality. Most authors conclude that there is an unsurmountable
barrier for the rms error (which, depending on the optimization
measure, is of the order of \unit[20]{\%}) resulting in a stalemate
when determining the ``best'' model. In this contribution, we have
shown that there are other criteria for assessing model quality, namely
completeness, robustness, parameter orthogonality, and intuitive
parameters that are calibrated to plausible values. For example, in
terms of the rms fitting error, the full-velocity difference model is
only marginally worse than the IDM or Gipps' model. Nevertheless, it
is significantly inferior in terms of our proposed criteria. 

Besides model quality, we also investigated the influence of the
calibration method and data
quality on the calibration result. It turned out that, generally, a
global calibration based on the gaps is more reliable and robust
compared to a local calibration or compared to a global calibration based on
speeds or accelerations, in agreement with~\cite{punzo2005analysis}. Concerning the data, we found that the common sampling
rates of \unit[10]{Hz} are unneccessarily high and that \unit[1]{Hz}
suffices. Furthermore, in contrast to intuition, smoothing the data
had no significant influence
on the calibration quality while data completeness, and also a minimum
total time interval of the order of \unit[300]{s}, and the absence of
significant behavioural changes are crucial. 

In a following contribution, we will investigate whether these results
which are obtained  for rather
high-quality xFCD and for a very limited selection of NGSIM trajectory
data will carry over to a systematic investigation of the noisier NGSIM trajectory
data. Further research fields concern the influence of inter- and
intra-driver variations. Regarding intra-driver variations, behavioural changes in
anticipation of or after actions not contained in the
models (such as lane changes) are highly relevant and deserve further
investigation.  It has been observed that 
different data sets can be described best by
different models representing different driving styles. We plan to
quantify this by proposing a model similarity matrix.

We conclude this contribution 
with some hints for performing a specific
calibration task:
\bi
\item Verify that the data do not describe traffic flow near a
``tipping point'' (e.g., at the verge of congestion or containing the
onset of a traffic breakdown) which is unsuitable for calibration.
\item Test the data for obvious behavioural changes in preparation for,
  or after, actions not desribed by the acceleration model, e.g.,
  active or passive  lane changes. Eliminate these time intervals or
  use a model including such actions.
\item Select a model with intuitive parameters and plausible
(published) values.
\item Identify which parameters of a model are relevant for 
the traffic situations to be found in the data. Keep the other
parameters fixed at published or plausible values.
\item Restrict the remaining parameter space by a bounding box
containing all plausible parameter
combinations. At any stage of estimation, the space outside is off-limits.
\item Take care to find a good initial guess. If in doubt, start with
the published values.
\item Check the resulting estimate for plausibility. Plot the fitness landscape
around the estimate to verify that there is a global minimum inside
the bounding box.
\ei

\section*{References}


\end{document}